\documentclass[%
 reprint,
 amsmath,amssymb,
 aps,
pra,
floatfix,
]{revtex4-1}

\usepackage{graphicx}
\usepackage{dcolumn}
\usepackage{bm}
\usepackage{amsfonts}
\usepackage{subeqnarray}
\usepackage{mathrsfs}
\usepackage{bbold}
\usepackage{tikz}
\usetikzlibrary{arrows}
\usepackage[caption=false]{subfig}
\newcommand{\angstrom}{\text{\normalfont\AA}}

\begin{document}

\title{Nonuniform superconductivity in wires with strong spin-orbit coupling}
\author{J. Baumard$^{1,2}$}
\author{J. Cayssol$^1$}%
\author{A. Buzdin$^{1}$}%
\affiliation{$^1$Univ. Bordeaux, CNRS, LOMA, UMR 5798, F-33405 Talence, France}
\affiliation{$^2$Donostia International Physics Center (DIPC), Manuel de Lardizabal 5, E-20018 San Sebasti\'an, Spain}


\begin{abstract}
We study theoretically the onset of nonuniform superconductivity in a one-dimensional single wire in presence of Zeeman (or exchange field) and spin-orbit coupling. Using the Green's function formalism, we show that the spin-orbit coupling stabilizes modulated superconductivity in a broad range of temperatures and Zeeman fields. We investigate the anisotropy of the temperature-Zeeman field phase diagram, which is related to the orientation of the Zeeman field. In particular, the inhomogeneous superconducting state disappears if this latter field is aligned or perpendicular to the wire direction. We identify two regimes corresponding to weak and strong spin-orbit coupling respectively. The wave-vector of the modulated phase is evaluated in both regimes. The results also pertain for quasi-1D superconductors made of weakly coupled 1D chains. 
\end{abstract}
\maketitle

\section{\label{sec:introduction}Introduction}
Superconducting systems in the presence of Zeeman field and spin-orbit coupling (SOC) exhibit  striking spectral and transport  properties including modifications in the pairing correlations \cite{loder_superconductivity_2013,zwicknagl_critical_2017}, the presence of Majorana zero-modes \cite{reeg_transport_2017,zha_majorana_2015,oreg_helical_2010,alicea_new_2012,beenakker_search_2013,krogstrup_epitaxy_2015,Sestoft:2018,shabani:2016}, unconventional magnetoelectric effect \cite{jacobsen_critical_2015,konschelle_theory_2015,mironov_spontaneous_2017,ojanen_magnetoelectric_2012,bobkova_gauge_2017,robinson:2019},  or critical field enhancement \cite{samokhin_magnetic_2004,barzykin_inhomogeneous_2002}. This latter feature was first observed in the bulk non-centrosymmetric heavy fermion superconductor CePt$_3$Si \cite{bauer_heavy_2004} and shortly after in CeRhSi$_3$ \cite{kimura_pressure-induced_2005} and CeIrSi$_3$ \cite{sugitani_pressure-induced_2006}.
Moreover, it has been predicted that the interplay between an in-plane magnetic field and SOC in a surface superconductor would lead to an inhomogeneous superconducting phase \cite{barzykin_inhomogeneous_2002,dimitrova_theory_2007,agterberg_magnetic-field-induced_2007}, similar to the  Fulde-Ferrell-Larkin-Ovchinnikov (FFLO) phase  \cite{fulde_superconductivity_1964,larkin_nonuniform_1964}. Low dimensional systems were recently fabricated using strong spin-orbit coupled semiconducting epitaxial InAsSb nanowires coated by an extremely thin layer of superconducting aluminium \cite{krogstrup_epitaxy_2015,Sestoft:2018} or in 2D platforms suitable for manipulating Majorana end states \cite{shabani:2016}. Besides the Majorana physics, such low-dimensional systems should also display a superconducting modulated phase. Moreover, such a modulated phase is characterized by the rotation of the superconducting phase along the wire, which produces a ground state with a finite phase difference $\varphi_0$ for a finite length wire used as a weak link \cite{krive_influence_2005, buzdin_direct_2008, reynoso_anomalous_2008,zazunov:2009}. Recently such $\varphi_0$-Josephson junction has been realized experimentally in nanowire quantum dots \cite{szombati_josephson_2016}.

The inhomogeneous FFLO state originates from the interaction of the superconducting condensate with a Zeeman field (see \cite{matsuda_fulde-ferrell-larkin-ovchinnikov_2007} as a review). The latter causes a spin-splitting, which in turn leads to Cooper pairs with a  finite total  momentum.  For certain values of the Zeeman field and temperature, the ground state has an oscillating  superconducting order parameter: $\Delta(\vec{r}) \propto e^{\text{i}\vec{q}\cdot\vec{r}}$, where $\vec{q}$ is the total momentum of the Cooper pairs.  As shown in Fig.  \ref{FFLO-diag}, in the absence of spin-orbit coupling, the FFLO state appears at low temperatures and high magnetic field. 
The transition between the normal and the FFLO states is of second order \cite{buzdin_nonuniform_1987,buzdin_generalized_1997} in the ballistic limit.

In a superconducting system, the interplay between Zeeman field $\vec{h}$ and spin-orbit coupling leads to the following additional term in the Ginzburg-Landau functional \cite{edelstein_ginzburg_1996}: 
\begin{equation}
-\text{i}\,\varepsilon\,\vec{n}\cdot\left[\vec{h}\wedge\left(\Delta^{\star}\vec{\nabla}\Delta - \Delta\vec{\nabla}\Delta^{\star}\right)\right]  \, ,
\label{GL}
 \end{equation}
where $\vec{n}$ is a unit vector along the asymmetric potential gradient and the phenomelological constant $\varepsilon$ is proportional to the microscopic spin-orbit coupling constant $\alpha$. In the following, we will use the constant $\alpha$, defined below in Eq. (\ref{H}), to characterize the spin orbit coupling strength. The term Eq. (\ref{GL}) is always non-zero if the direction of the field $\vec{h}$ does not coincide with the direction of the spin-orbit vector $\vec{n}$ and may result in the formation of the modulated state with finite $\vec{q}$. This has been theoretically studied in several works on infinite 2D and 3D systems \cite{mineev_helical_1994,loder_superconductivity_2013,dimitrova_theory_2007,houzet_quasiclassical_2015, zwicknagl_critical_2017, agterberg_magnetic-field-induced_2007, barzykin_inhomogeneous_2002, kaur_helical_2005}. 

In this paper, we investigate the formation of inhomogeneous superconductivity in the presence of spin-orbit interaction and Zeeman (or exchange) field $\vec{h}$ for the particular situation of one-dimensional systems. We consider a superconducting wire deposited on a substrate (Fig. \ref{schema_system}). In this geometry, the modulation may occur only along the wire direction $x$, and the vector $\vec{n}$ is normal to the substrate (along $y$). This implies that the field $\vec{h}$ must have a component along the third direction, here $z$, in order to observe the effects related to the coupling Eq. (\ref{GL}). Moreover, in the case of the exactly quadratic electron spectrum, 1D systems have a peculiar behavior: the modulated phase also requires a non zero component along the wire direction \cite{mironov_double_2015}. Consequently, the component $h_x$ of the field is also required to generate the nonuniform phase \cite{nesterov_anomalous_2016,ojanen_magnetoelectric_2012}. Using the Green's function formalism, we provide a self-consistent study of the effect of the magnetic field orientation on the field-temperature phase diagram of 1D systems both in the weak and strong spin-orbit coupling regimes. Our model is also relevant for quasi-1D superconductors like some organic superconductors \cite{buzdin_organic_1984,buzdin_nonuniform_1987,croitoru_peculiarities_2014}. Such compounds consist of weakly coupled 1D chains. The interchain coupling is described by a hopping parameter $t$. If the hopping integral $t \ll T_c$, then the system can be described by a strictly 1D model. On the other hand, as it has been demonstrated in \cite{tsuzuki_long-range_1972}, the mean-field treatment is justified if $t \gg T_c^2/E_\text{F}$. So for $T_c^2/E_\text{F} \ll t \ll T_c$, the critical fluctuations of the superconducting order parameter are effectively suppressed, and the system can be treated as a strictly superconducting 1D wire. 

The paper is organized as follows. In section \ref{sec:weak_soc}, we derive the Gor'kov equations \cite{gorkov_energy_1958,abrikosov_methods_1965} for our system, and solve them to obtain the anomalous Gor'kov Green's function Eq. (\ref{eq. Fdu-single}) describing superconducting correlations for arbitrary exchange field and spin-orbit coupling. This solution is used to write the self-consistency gap equation in the regime of small spin-orbit fields $\alpha\,p_\text{F} \ll T_c$, and demonstrate the presence of a modulated superconducting phase. The modulation wavevector scales as $h_x^2\,h_z\,\alpha^3$. In section \ref{sec:strong_soc}, we consider the regime of large spin-orbit coupling $\alpha\,p_\text{F} \gg T_c$. In the strong spin-orbit coupling regime, the electronic spectrum is drastically modified, and it is necessary to work in the so-called helical basis to describe the normal state. We further derive the self-consistency relation in terms of the helical Green's functions. Analysis of the self-consistency relation reveals that increasing the ratio $\alpha\,p_\text{F}/h$ leads to change the nature of superconducting correlations from interband to intraband. We also find an inhomogeneous superconducting phase in this regime, with a wave vector of the modulation scaling as $h_x^2\,h_z\,/\alpha$, for large values of $\alpha$. We finally provide a complete $(h_x,\, h_z,\, T)$-phase diagram for different orientations of the Zeeman field and several values of the spin-orbit coupling constant $\alpha$.

\section{\label{sec:weak_soc} Weak spin-orbit coupling}
We consider a single 1D superconducting wire deposited on top of a substrate (Fig. \ref{schema_system}). The mirror symmetry breaking induces Rashba spin-orbit interaction in the wire. A magnetic field can be applied in the plane $x-z$ corresponding to the surface of the substrate, thereby generating the Zeeman coupling. After introducing the Hamiltonian, we derive the corresponding exact Gor'kov equations for the Green's functions of the 1D wire. The self-consistency relation is then studied in the limit of weak spin-orbit interaction and small wave-vector $q$ of the modulated superconducting order parameter.  

\subsection{\label{subsec:model_weak} Model and Hamiltonian}
\begin{figure}
\includegraphics[scale=0.43]{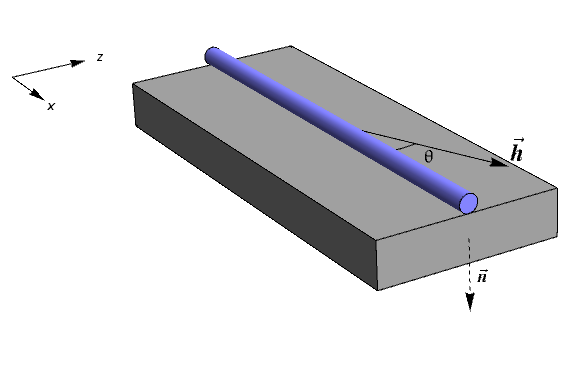}
\caption{We consider an infinite superconducting wire along the $x$-axis. A magnetic field is applied within the plane $x-z$ corresponding to the surface of the substrate, leading to a Zeeman coupling where $\vec{h}=(h_x,\,0,\,h_z)$. Inversion symmetry is broken along the direction of the unit vector $\vec{n}$ normal to the substrate surface, thus inducing Rashba spin-orbit interaction in the wire. \label{schema_system}}
\end{figure}

Let us consider a superconducting wire deposited on top of a substrate (Fig. \ref{schema_system}). The wire is assumed to be infinite along the $x$-axis, and a Zeeman field $\vec{h}=(h_x,\,0,\,h_z)$ is applied within the plane $x-z$, thereby making an arbitrary angle $\theta$ with the wire. 
The second quantized Hamiltonian of the 1D wire is expressed as:  
\begin{equation}
\mathcal{H} = \mathcal{H}_{0} + \mathcal{H}_{\vec{h}} + \mathcal{H}_\text{SO} + \mathcal{H}_\text{BCS}\; ,
\label{Hamiltonian1}
\end{equation}
where 
\begin{subequations}
\begin{align}
\mathcal{H}_{0} &= \displaystyle \sum_{\rho} \int \mathrm{d}x \, \psi^{\dagger}_{\rho}\,\left(\frac{-\hbar^2}{2\,m}\,\partial^2_x - \mu \right)\,\psi_{\rho}\; , \\
\mathcal{H}_{\vec{h}} &= \displaystyle \sum_{\rho,\, \rho^{\prime}} \int \mathrm{d}x \, \psi^{\dagger}_{\rho}\,\left[\vec{h}\cdot\vec{\sigma}\right]_{\rho\,\rho^{\prime}}\,\psi_{\rho^{\prime}}\; ,\\
\mathcal{H}_\text{SO} &= 
\alpha\,\frac{\hbar}{\text{i}}\,\displaystyle \sum_{\rho,\, \rho^{\prime}} \int \mathrm{d}x \, \psi^{\dagger}_{\rho}\,\left[\sigma_z\right]_{\rho\,\rho^{\prime}}\,\partial_x\,\psi_{\rho^{\prime}}\; ,\\
\mathcal{H}_\text{BCS} &= \displaystyle \int \mathrm{d}x \left(\Delta(x)\,\psi^{\dagger}_{\uparrow}\,\psi^{\dagger}_{\downarrow} + \Delta^{\star}(x)\,\psi_{\downarrow}\,\psi_{\uparrow}\right)\;.
\end{align}
\label{H}
\end{subequations}
The annihilation electronic field operator at position $x$ is written $\psi_\rho=\psi_\rho(x)$, where $\rho$ is the spin index. The summation is made over the spin $\rho$, and the Pauli matrices $\sigma_i$ ($i = x, \, y,\, z$) have the standard forms :
\begin{center}
$\sigma_x = \begin{pmatrix}
0 & 1 \\ 1 & 0
\end{pmatrix}$\;;
$\sigma_y = \begin{pmatrix}
0 & -\text{i} \\ \text{i} & 0
\end{pmatrix}$\;;
$\sigma_z = \begin{pmatrix}
1 & 0 \\ 0 & -1
\end{pmatrix}$\;.
\end{center}

The kinetic term $\mathcal{H}_0$ corresponds to a purely quadratic dispersion with an effective mass $m$ and a chemical potential $\mu$. The Zeeman Hamiltonian $\mathcal{H}_{\vec{h}}$ incorporates the effect of a tilted field with respect to the wire direction $x$. The Rashba Hamiltonian $\mathcal{H}_\text{SO}$ originates from the mirror symmetry breaking due to the substrate, and its strength is given by the spin-orbit velocity $\alpha$. Finally, $\mathcal{H}_\text{BCS}$ encodes the standard s-wave superconducting correlations within the BCS model.

Because of the Zeeman splitting, the singlet superconducting order parameter can oscillate along the $x$-axis as:
\begin{equation}
\Delta(x) = \Delta_0\,e^{\text{i} q x}\;,
\label{Delta}
\end{equation}
where $\Delta_0$ is independent of the position $x$, and $q$ is the wave vector of the inhomogeneous order parameter.

\subsection{\label{subsec:Gorkov_eq} Gor'kov equations}
The Gor'kov Green's functions are defined as the following $2$-point correlators between the field operators:
\begin{eqnarray}
G_{\alpha \beta}(x,x',\tau) &=& - \left\langle T_{\tau}\,\psi_{\alpha}(x,\tau)\,\psi^{\dagger}_{\beta}(x',0)\right\rangle \;,\label{G}\\
F^{\dagger}_{\alpha \beta}(x,x',\tau) &=& \left\langle T_{\tau}\,\psi^{\dagger}_{\alpha}(x,\tau)\,\psi^{\dagger}_{\beta}(x',0)\right\rangle \;,\label{Fdag}
\end{eqnarray}
where $T_\tau$ is the time-ordering product for the imaginary time $\tau$, and the brackets denote averaging over the thermal equilibrium distribution at temperature $T$ and chemical potential $\mu$. The electronic field operators obeys the fermionic anticommutation relations at equal (imaginary) times.

We follow the standard procedure \cite{abrikosov_methods_1965, kopnin_theory_2001} to establish the differential equations for the Gor'kov Green's functions. First, the equations of motion for the field operators $\psi_{\alpha}(x,\tau)$ and $\psi_{\alpha}^\dagger(x,\tau)$  are obtained by evaluating their commutators with the Hamiltonian Eq. (\ref{H}). Second, multiplying the latter equations of motion by the field operators at $(x',0)$ and averaging other the thermal distribution leads to relations between the correlators Eqs. (\ref{G},\ref{Fdag}) and their $\partial_\tau$ derivatives with respect to imaginary time. Due to invariance with respect to time translations in $\tau$, it is then convenient to express those equations in Fourier space of Matsubara frequencies $\omega_n$ :
\begin{subeqnarray}
\left( \text{i}\,\omega_n + M(x)\right)\,G + \Delta(x) \, \text{i}\,\sigma_y\,F^{\dagger} &=& \delta(x-x^{\prime})\,\mathbb{1} \; ,\slabel{Gorkov1}\\
\left( \text{i}\,\omega_n - M^{\star}(x)\right)\,F^{\dagger} - \Delta^{\star}(x)\,\text{i}\,\sigma_y\,G &=& 0 \;,
\slabel{Gorkov2}
\end{subeqnarray}
where $G =G(x,x',\omega_n) $ and $F =F(x,x',\omega_n) $ are two by two matrices defined in spin space, and $\omega_n = 2\,\pi\,k_\text{B}\,T\,\left(n + 1/2 \right)$, $n$ being integer. The differential operator $M(x)$ is defined by :
\begin{equation}
M(x) =  \frac{\hbar^2}{2\,m}\,\partial^2_x + \mu - \vec{h}\cdot\vec{\sigma} - \alpha\,\sigma_z\,\frac{\hbar}{\text{i}}\,\partial_x \; ,
\end{equation}
where $\mathbb{1}$ corresponds to the identity matrix. To simplify the expressions, we will take $\hbar = k_\text{B} = 1$ in the following.

\subsection{\label{subsec:Fdu_weak} Solution near the second order transition}
Here we solve the Gor'kov equations near the second order phase transition between the normal and superconducting states, namely assuming the limit $\Delta_0 \ll T_{c0}$. 

To this purpose, it is useful to introduce the function $G_0$ which solves Eq. (\ref{Gorkov1}) for $\Delta_0=0$, and also a closely related function $g_0$ which is the opposite of the kernel appearing in Eq. (\ref{Gorkov2}). Those functions $G_0$ and $g_0$ are defined by the equations :
\begin{eqnarray}
\left[\text{i}\,\omega_n + M(x) \right] G_0 &=& \delta(x-x^{\prime})\,\mathbb{1}\; ,\\
-\left[ \text{i}\,\omega_n - M^{\star}(x)\right] g_0 &=& \delta(x-x^{\prime})\,\mathbb{1}\, ,
\end{eqnarray}
which are easily solved in momentum space, yielding :
\begin{equation}
G_0 = \frac{1}{D_-}
\begin{pmatrix}
\text{i}\,\omega_n - \xi + \alpha\,p + h_z & h_x \\
h_x & \text{i}\,\omega_n - \xi - \alpha\,p - h_z
\end{pmatrix}\;,
\label{G0}
\end{equation}
and:
\begin{equation}
g_0 = -\frac{1}{D_+}
\begin{pmatrix}
\text{i}\,\omega_n + \xi + \alpha\,p - h_z & -h_x \\
-h_x & \text{i}\,\omega_n + \xi - \alpha\,p + h_z
\end{pmatrix}\;,
\label{g0}
\end{equation}
where $\xi = \xi_p = \frac{p^2}{2\,m} - \mu$, and the $(p,\omega_n)$ arguments have been omitted in $G_0(p)=G_0(p,\omega_n)$ and $g_0(p)=g_0(p,\omega_n)$. The denominators $D_\pm$ are given by :
\begin{equation}
D_{\pm} = \left[\text{i}\,\omega_n \pm \left(\xi + \alpha\,p\right) - h_z\right]\left[\text{i}\,\omega_n \pm \left(\xi - \alpha\,p\right) + h_z\right] - h_x^2\; .
\label{D}
\end{equation}

In the limit $\Delta_0 \rightarrow 0$, the Gor'kov Green's function $F^\dagger$ can be written as :
\begin{equation}
F^{\dagger} = -\Delta_0\,g_0(p)\,\text{i}\,\sigma_y G_0(p + q)\; .
\label{F}
\end{equation}
Finally, using Eqs. (\ref{G0}, \ref{g0}, \ref{D}) to compute the off-diagonal entry $F_{\downarrow\uparrow}^{\dagger}$ of Eq. (\ref{F}), one obtains : 
\begin{widetext}
\begin{equation}
\frac{F_{\downarrow\uparrow}^{\dagger}}{\Delta_0} = -\frac{\left(\text{i}\,\omega + \xi_p - \alpha\,p + h_z\right)\,
\left(i\,\omega - \xi_{p+q}+\alpha\left(p+q\right) + h_z\right) + h_x^2}{\left[\left(\text{i}\,\omega + \xi_p + \alpha\,p - h_z\right)\,
\left(\text{i}\,\omega + \xi_p - \alpha\,p + h_z\right) - h_x^2\right]\left[\left(\text{i}\,\omega - \xi_{p+q}-\alpha\left(p+q\right) - h_z\right)\,
\left(\text{i}\,\omega - \xi_{p+q}+\alpha\left(p+q\right) + h_z\right) - h_x^2\right]}\;.
\label{eq. Fdu-single}
\end{equation}
\end{widetext}
This expression Eq. (\ref{eq. Fdu-single}) is the solution of the linearized Gor'kov equations for the anomalous Green's function. At this stage, no assumptions are made upon the Zeeman fields ($h_x$ and $h_z$) and spin-orbit coupling $\alpha$ in the above equation, the only assumption being the limit $\Delta_0 \rightarrow 0$. 

\subsection{\label{subsec:DL_weak} Self-consistency relation and critical temperature}

The self-consistency equation for the superconducting order parameter, $\Delta(x)= |\gamma| F_{\downarrow\uparrow}^{\dagger}(x,x,\tau=0) $,  reads in momentum and Matsubara frequency representation :
\begin{equation}
\Delta_0 = |\gamma|\,T\,\displaystyle\sum_{\omega_n} \int_{-\infty}^{+\infty}F_{\downarrow\uparrow}^{\dagger}(p)\frac{\mathrm{d}p}{2\,\pi}\;,
\label{SCE1}
\end{equation}
where $\gamma$ is the attractive electron-electron BCS effective coupling constant. Near the critical temperature, Eq. (\ref{SCE1}) can be written in a more convenient form, eliminating $\gamma$ :
\begin{equation}
\ln \left( \frac{T_c}{T_{c0}} \right) = 2\, T_c\, \sum_{\omega_n\,\geq\,0}\left[\frac{v_\text{F}}{2}\,\int_{-\infty}^{+\infty}\mathrm{Re}\left(\frac{F^{\dagger}_{\downarrow\uparrow}}{\Delta_{0}}\right)\,\mathrm{d}p -  \frac{\pi}{\omega_n} \right]\;,
\label{eq. SC}
\end{equation}
where $v_\text{F}$ is the Fermi velocity in the absence of SOC, $T_c$ is the critical temperature of the wire in presence of Zeeman and Rashba interactions, while $T_{c0}$ is the critical temperature in the absence of these fields, i.e. for $h_{x}=h_{z}=0$ and $\alpha=0$. 

A simple way of seeing the influence of spin-orbit coupling on the emergence of the modulated phase is to expand Eq. (\ref{eq. SC}) in series with respect to the wave-vector $q$ as :
\begin{eqnarray}
\ln\left(\frac{T_c}{T_{c0}}\right) &=& 2\,\pi\,T_c\,\sum_{\omega_n \geq 0}\left[-\frac{h_x^2 + h_z^2}{\omega_n^3} + \frac{7\,p_\text{F}^2}{\omega_n^7}\,h_x^2\,h_z\,\alpha^3\,q\right.\notag\\ 
&&\left.- \frac{v_\text{F}^2}{4\,\omega_n^3}\,q^2\right]\;.
\label{SCE-DL}
\end{eqnarray}
This expansion is justified by the fact that the wave-vector $q$ is small near the transition towards the nonuniform phase. We derived Eq. (\ref{SCE-DL}) assuming a weak Zeeman field ($h_x,\, h_z \ll T_c$) and weak spin-orbit coupling ($\alpha\,p_\text{F} \ll T_c$). The chemical potential is larger than all the other energies: $\mu \gg h_x,\,h_z,\, E_\text{so}$, where $E_{so} = \frac{1}{2}\,m\,\alpha^2$. Therefore the Green's function (\ref{eq. Fdu-single}) may be expanded in series with respect to the small energies $h_x$, $h_z$ and $\alpha\,p$. Because of the $p$ integration in Eq. (\ref{eq. SC}), only even in momentum terms will remain, allowing to transform the integral over $p$ into an integral over $\xi$ following:
\begin{equation}
\int_{-\infty}^{+\infty} \mathrm{d}p =  2\,\int_{-\mu}^{+\infty} N\left(\xi\right) \mathrm{d}\xi  \;,
\label{eq:change_int}
\end{equation}
where the density of states at energy $\xi$ reads :
\begin{equation}
N\left(\xi\right) = \displaystyle\sqrt{\frac{m}{2\left(\xi + \mu\right)}} \, .
\end{equation}
For a finite spin-orbit coupling, the Green's function $F^\dagger_{\downarrow\uparrow}$ contains terms proportional to $p^2 = 2\,m\left(\xi + \mu\right)$ and $p^4$ in the numerator. Then, the presence of the small parameters $\xi/\mu$ and $\xi^2/\mu^2$ in $G$ implies that we should expand $N\left(\xi\right)$ at the second order in $\xi/\mu$. This is a specificity of spin-coupled systems. In the absence of SOC, it is sufficient to keep the density of state $N\left(\xi\right)$ at zero order in $\xi/\mu$ to compute the Green functions. Finally, the integral Eq. ($\ref{eq:change_int}$) becomes:
\begin{equation}
\int_{-\infty}^{+\infty} \mathrm{d}p \rightarrow \frac{2}{v_\text{F}} \int_{-\infty}^{+\infty} \left(1 - \frac{\xi}{2\,\mu} + \frac{3\,\xi^2}{8\,\mu^2}\right) \mathrm{d}\xi  \;,
\label{eq:int_xi}
\end{equation}
where $v_\text{F} = \displaystyle\sqrt{2\,m/\mu}$. The $\xi$ integration is then performed using the residue technique.

The main result is the presence of a linear term in $q$ which leads to the emergence of a modulated superconducting state at wave-vector $q_0$ given by :
\begin{equation}
    \hbar q_0   = \frac{127\,m^2\,h_x^2\,h_z\,\alpha^3}{8\,\pi^4\,T_c^4}\,\frac{\zeta(7)}{\zeta(3)} \;,
    \label{q0_weak}
\end{equation}
where the factor $\hbar$ has been restored (previously $\hbar=1$ in the derivation of the this result). The Euler-Riemann zeta function is denoted $\zeta(s)$.
Hence the modulation requires spin-orbit coupling and a tilted Zeeman field, with $h_x$ and $h_z$ both finite, and $q_0 \propto h^3\,\alpha^3$, which is in contrast with the 2D case, where $q_0 \propto h\,\alpha^3$ \cite{konschelle_theory_2015, edelstein_magnetoelectric_1995}. 

\begin{figure*}
\begin{center}
\subfloat[\label{FFLO-diag}]{\includegraphics[scale=0.64]{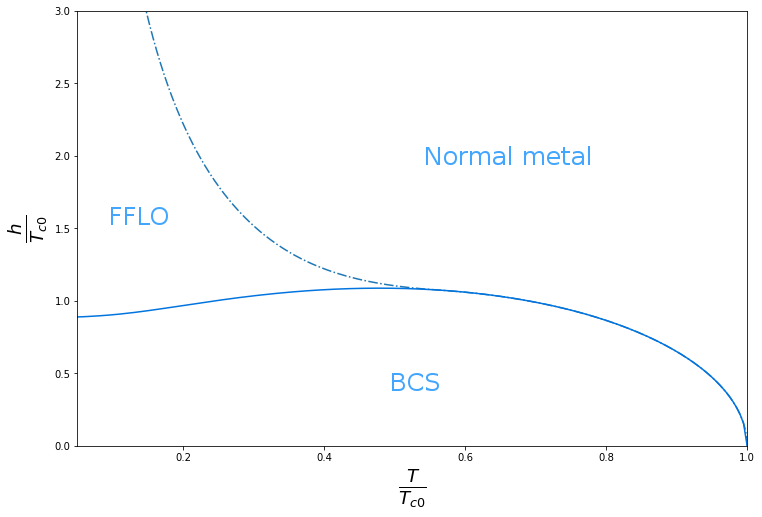}}
\hspace{0.5cm}
\subfloat[\label{diag-theta}]{\includegraphics[scale=0.32]{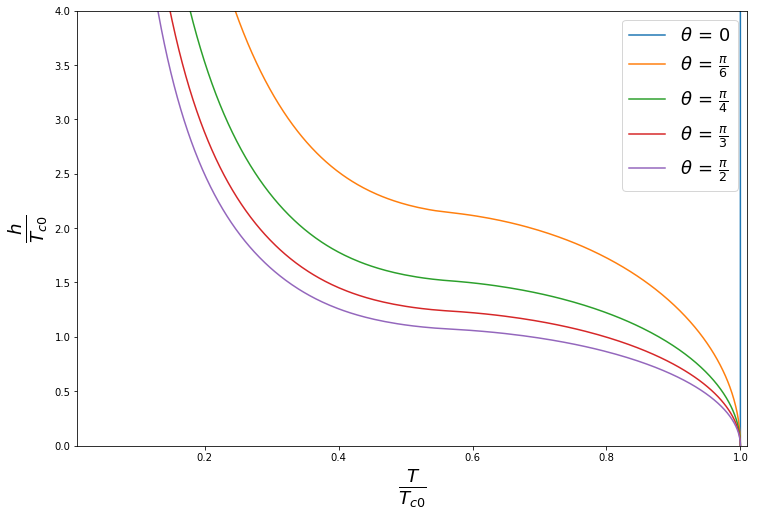}}\\
\end{center}
\caption{Phase diagrams in the Zeeman field-temperature $(h,T)$ plane, for a one-dimensional wire with Fermi energy $\mu = 10^3\,T_{c0}$. \protect\subref{FFLO-diag} In the absence of spin-orbit coupling ($\alpha=0$), the FFLO phase diagram is characterized by a tricritical point located at coordinates $T^\star = 0.56\,T_{c0}$ and $h^{\star} = 1.07\,T_{c0}$. It corresponds to the meeting point of three transition lines, which separate the normal metal, the uniform superconductor (BCS) and nonuniform FFLO state. \protect\subref{diag-theta} In the presence of spin-orbit interaction with strength $\alpha = 0.05\,v_\text{F}$ ($\alpha\,p_\text{F} = 100\,T_{c0}$), transition lines between the normal and the superconducting states for different orientations of the Zeeman field. The modulus of the field is denoted by $h = |\vec{h}|$, while $\theta$ corresponds to the angle between $\vec{h}$ and the wire, see Fig. \ref{schema_system}. \label{diag_phase}}
\end{figure*}

We conclude this section by a discussion of the observability of the modulated state in wires with weak SO coupling. In order to evaluate the order of magnitude of $q_0$, we replace the mass $m$ in Eq. (\ref{q0_weak}) by its expression in terms of $\mu$ and $k_F$ for an ideal parabolic band. We obtain :
\begin{equation}
     q_0  \simeq  \frac{127}{32\,\pi^4} \frac{\zeta(7)}{\zeta(3)}  \, \, \frac{h_x^2\,h_z\,\tilde{\alpha}^3}{\,T_c^4 \mu^2}  k_F^4 \, \,\;,
    \label{q0_weak_eval}
\end{equation}
where the $\hbar$ factors disappears. The spin-orbit coupling has been expressed by the parameter $ \tilde{\alpha}$ :
\begin{equation}
    \tilde{\alpha}= \hbar \alpha  \,\;,
    \label{alphatilde}
\end{equation}
which is commonly used in experimental works on spin-orbit systems. The modulation is very sensitive to the value of this spin-orbit coupling. In the regime of 
weak SO coupling corresponding to the range $ \tilde{\alpha}= 1-10 \, \,  {\rm meV}.\angstrom$, the modulation wavevector $q_0$ reaches extremely small values (in comparison to $k_F \simeq 0.6$ A$^{-1}$) which justifies {\it a posteriori} the expansion in $q$. For the other parameters, we have fixed the values  $h_x=h_z=0.5$ meV for the Zeeman components, $T_c = 1$ meV, and $\mu = 1$ eV for the Fermi energy. In order to obtain $q_0$ which could be experimentally observable, it is necessary to "push" the SO parameter to values such as $ \tilde{\alpha}= 100 \, \,  {\rm meV}.\angstrom$, which are now experimentally available but fall far beyond the domain of validity of the above result Eq. (\ref{q0_weak}). 
 
\section{\label{sec:strong_soc} Strong spin-orbit coupling}
In the previous section, the self-consistency relation was derived in the case of weak spin-orbit interaction: $\alpha\,p_\text{F} \ll T_c$. This section is devoted to the opposite situation characterized by a strong spin-orbit coupling $\alpha\,p_\text{F} \gg T_c$. After rewriting the normal Hamiltonian of Eq. (\ref{H}) in the \textit{helical basis} with diagonalizes it, we derive the self-consistency relation, in the limit of a small $\Delta_0$, but for strong spin-orbit coupling. This allows to plot the $\left(h_x,h_z,T\right)$-phase diagram for various orientations of the Zeeman field and large spin-orbit coupling constants. The limit of strong-spin orbit coupling, typically $ \tilde{\alpha}= 300 \, \,  {\rm meV.A}$, can be achieved in narrow gap semiconductors like InAs or InSb combined with aluminium as superconductor \cite{krogstrup_epitaxy_2015,Sestoft:2018,shabani:2016}.

\subsection{\label{subsec:helical_basis} Hamiltonian in the helical basis}
We still consider the same model Eq. (\ref{H}) for the wire, but we are going to use a different basis to facilitate the study of the large spin-orbit coupling regime. In the absence of superconductivity, the Hamiltonian describing the wire reads :
\begin{equation}
    \hat{h}_\text{N} = \left(\xi + \alpha\,p\,\sigma_z\right)\,\tau_z + \vec{h}\cdot\vec{\sigma} \, ,
\end{equation}
where $\sigma_{x,y,z}$ and $\tau_{x,y,z}$ are the Pauli matrices acting in spin and Nambu space respectively.

Using the unitary transformation 
\begin{equation}
   \hat{h}_\text{N} \rightarrow \mathcal{U}^\dagger\,\hat{h}_\text{N}\,\mathcal{U} = \hat{h}_\text{hel} \, ,
\end{equation}
with $ \mathcal{U}$ given explicitly in Appendix \ref{app:hamiltonian}, the normal state Hamiltonian can be written in the helical basis as :
\begin{equation}
    \hat{h}_\text{hel} = \xi\,\tau_z - \frac{\lambda}{2}\left[\left(\varepsilon_+ + \varepsilon_-\right)\,\tau_0 + \left(\varepsilon_+ - \varepsilon_-\right)\,\tau_z\right]\;,
    \label{hamiltonian_diag}
\end{equation}
where $\lambda = \pm$ labels the energy bands, and $\varepsilon_\pm = \sqrt{h_x^2 + \left(h_z \pm \alpha\,p\right)^2}$. Matrix $\tau_0$ is the identity matrix acting in Nambu space.

This allows to solve the equations for the normal state Green's functions for each helical band independently : 
\begin{equation}
\left(\text{i}\,\omega_n - \hat{h}_\text{hel}\right)\hat{G}_0 = \mathbb{1} \, ,
\end{equation}
which leads to the solutions in Nambu-helical space:
\begin{equation}
    \hat{G}_0 = \begin{pmatrix}
    G_\lambda\left(p,\, \omega_n\right) & 0\\
    0 & -G_\lambda\left(-p,\, -\omega_n\right)
    \end{pmatrix} \;,
\end{equation}
where
\begin{equation}
   G_\lambda\left(p,\, \omega_n\right) = \frac{1}{\text{i}\,\omega_n - \xi + \lambda\,\sqrt{h_x^2 + \left(h_z + \alpha\,p\right)^2}} \;.
\end{equation}

\begin{figure*}
\begin{center}
\includegraphics[scale=0.38]{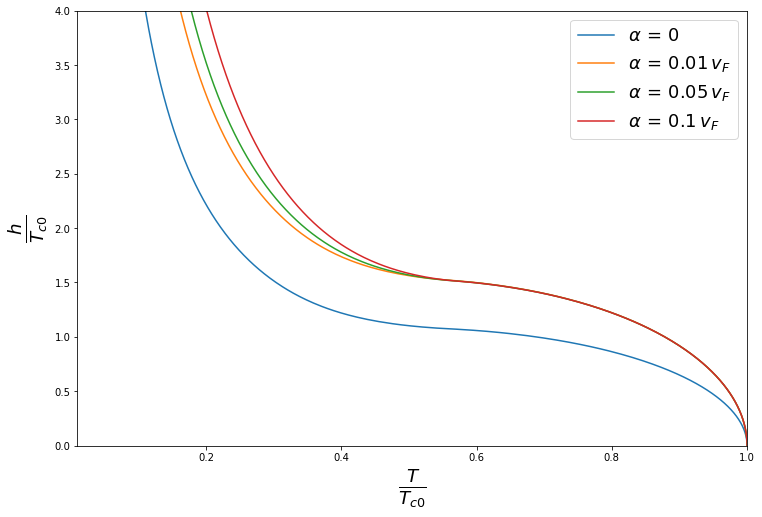}
\end{center}
\caption{Phase diagram in $(h,T)$ for different values of the spin-orbit coupling constant: $\alpha = 0$, $\alpha = 0.01\,v_\text{F}$ ($\alpha\,p_\text{F} = 20\,T_{c0}$), $\alpha = 0.05\,v_\text{F}$ ($\alpha\,p_\text{F} = 100\,T_{c0}$) and $\alpha = 0.1\,v_\text{F}$ ($\alpha\,p_\text{F} = 200\,T_{c0}$). The Zeeman field is taken such that $h_x = h_z$ ($\theta = \pi/4$). The Fermi energy is $\mu = 10^3\,T_{c0}$ \label{diag_phaseSO}}
\end{figure*}

Let us now consider the system in the presence of superconductivity in the limit $\Delta_0 \ll T_c$. In order to take into account the possibility of a modulated superconducting order parameter 
$\Delta_0 e^{iqx}$, the self-consistency equation reads
\begin{equation}
    \Delta_0 = \frac{|\gamma|}{4}\,T\,\sum_{\omega_n} \int_{-\infty}^{+\infty} \text{Tr}\left(\hat{G}^S\left(\mathcal{U}^{(q)}\right)^\dagger\tau_x\,\mathcal{U}^{(q)}\right) \frac{\mathrm{d}p}{2\,\pi} \;,
    \label{SCE_strong_SOC}
\end{equation}
where $\mathcal{U}^{(q)}$ is the "momentum shifted" version (Appendix \ref{app:Adding_SC}) of the unitary matrix connecting the original spin basis and the helical basis. The matrix $\hat{G}^S$ is the Green's function in the presence of superconductivity. At first order in $\Delta_0$:
\begin{equation}
    \hat{G}^S = \hat{G}_0^{(q)} - \Delta_0\hat{G}_0^{(q)}\left(\mathcal{U}^{(q)}\right)^\dagger\tau_x\,\mathcal{U}^{(q)}\,\hat{G}_0^{(q)} \;.
    \label{Gq}
\end{equation}
The label $S$ in $\hat{G}^{S}$ denotes the presence of superconductivity, and the superscript $q$ in $\hat{G}_0^{(q)}$ and $\mathcal{U}^{(q)}$ refer to the "shifted versions" of $\hat{G}_0$ and $\mathcal{U}$  defined in the appendix B. The Green's function $\hat{G}^S$ contains both intraband and interband terms (see Appendix \ref{app:SCE}). In the regime of strong spin-orbit interaction, $\alpha\,p_\text{F} \gg h_x,\,h_z,\,T_c$ and $E_\text{so} = \frac{1}{2}\,m\,\alpha^2 \ll \mu$, the interband terms are negligible and it is sufficient to consider only intraband pairing. It is convenient to eliminate $\gamma$, and rewrite the self-consistency relation Eq. (\ref{SCE_strong_SOC}) as :
\begin{widetext}
\begin{equation}
    \ln\left(\frac{T_c}{T_{c0}}\right) = T_c\,\sum_{\omega_n \geq 0,\,\lambda} \left[\frac{v_\text{F}}{2}\,\text{Re}\left(\int_{-\infty}^{+\infty} G_\lambda\left(p + \frac{q}{2},\, \omega_n\right)\,G_\lambda\left(-p + \frac{q}{2},\, -\omega_n\right)\mathrm{d}p\right) - \frac{\pi}{\omega_n}\right]\; ,
    \label{SCE_strong_SOC2}
\end{equation}
\end{widetext}
which is reminiscent of Eq. (\ref{eq. SC}) albeit expressed in the helical basis labelled by $\lambda$, rather than in the spin basis. For each helical band $\lambda$, the integral over momentum can be turned into an integral over $\xi$ using quasi-classical approach: 
\begin{equation}
    \int_{-\infty}^{+\infty} \mathrm{d}p \rightarrow \frac{1}{v_\text{F}}\,\int_{-\infty}^{+\infty} \mathrm{d}\xi \;.
\end{equation}
The integration over $\xi$ is then straightforward and the expression of the self-consistency equation is provided in Appendix \ref{app:SCE}.

\subsection{\label{subsec:diag} Modulated phase in the presence of strong SOC}

In the same way as Sec. \ref{subsec:DL_weak}, we first investigate the emergence of the modulated phase. We consider small values of the Zeeman field $h_x,\, h_z \ll T_c$ and small values of $q$ such that $q \ll p_\text{F}$:
\begin{equation}
    \ln\left(\frac{T_c}{T_{c0}}\right) = 2\,\pi\,T_c\,\sum_{\omega_n \geq 0}\left[-\frac{h_z^2}{\omega_n^3} + \frac{3\,h_x^2\,h_z}{2\,\omega_n^3\,\alpha\,p_\text{F}^2}\,q - \frac{v_\text{F}^2}{4\,\omega_n^3}\,q^2\right] \;.
    \label{DL_strong_SOC}
\end{equation}
The self-consistency relation Eq. (\ref{DL_strong_SOC}) was obtained by considering only intraband pairing. In this case, we can notice that when $h_z = 0$, Eq. (\ref{DL_strong_SOC}) leads to $T_c = T_{c0}$, which seems to indicate that superconductivity is not affected by the Zeeman field when this one is normal to the spin-orbit interaction. However, we have to keep in mind that taking interband correlations into account in Eq. (\ref{Trace}) would modify this result by terms of order $h_x^2/\alpha^2\,p_\text{F}^2$ in Eq. (\ref{DL_strong_SOC}).

The $q$ expansion of the self-consistency relation Eq. (\ref{DL_strong_SOC}) exhibits a linear term in $q$ which leads to the emergence of an inhomogeneous superconducting state at wave-vector:
\begin{equation}
    \hbar q_0 = \frac{3\,h_x^2\,h_z}{4\,\alpha\,\mu^2}\; ,
    \label{q0_strong}
\end{equation}
where $q_0$ is estimated by maximizing Eq. (\ref{DL_strong_SOC}) with respect to $q$. This expression of $q_0$ is only valid for large values of $\alpha$, and we have restored the factor $\hbar$. The scaling in $h_x$ and $h_z$ is similar to the one for weak spin orbit, see Eq. (\ref{q0_weak}). 

Introducing the parameter $\tilde{\alpha}$ commonly used in the experimental literature, this expressions reads :
\begin{equation}
     q_0 = \frac{3\,h_x^2\,h_z}{4\,\tilde{\alpha}\,\mu^2}\; .
    \label{q0_strong_eval}
\end{equation}
For the values $\tilde{\alpha} = 300$ meV.A of the SO coupling, $h_x=h_z=0.5$ meV of the Zeeman components, and $\mu = 1$ eV for the Fermi energy, this formula leads to a very small $q_0$ which justifies {\it a posteriori} the expansion in $q$. 

\subsection{\label{subsec:numerics} Numerical results and phase diagrams}

We now solve the self-consistency relation Eq. (\ref{app:SCE_fin}) for arbitrary values of the wave-vector $q$, i.e. without doing any expansion in parameter $q$. We solve numerically the self-consistency equation Eq. (\ref{app:SCE_fin}) in the following way. First, the temperature is extracted from Eq. (\ref{app:SCE_fin}) as a function of both Zeeman field $h$ and wave-vector $q$. The optimal wave-vector $q$ is then determined such that it maximizes the temperature at a given $h$, which is equivalent to minimize the Ginzburg-Landau free energy. 

\medskip

Following this procedure, the $(h_x,\, h_z,\, T)$-phase diagram is plotted (Fig. \ref{diag-theta}) for different orientations of the field and for spin-orbit coupling  $\alpha = 0.05\, v_\text{F}$ ($\alpha\,p_\text{F} = 100\,T_{c0}$). One can first observe the anisotropy induced by spin-orbit interaction: 	at fixed critical field (temperature), the critical temperature (field) increases when $\theta$ decreases, $\theta$ corresponding to the angle between the field and the wire (see Fig. \ref{schema_system}). This means that the superconducting phase is widened when the field tends to be normal to the SOC. As expected from Eq. (\ref{DL_strong_SOC}), the purely parallel Zeeman field $h = h_z$ ($\theta = \pi/2$) or purely perpendicular $h = h_x$ ($\theta = 0$) to the spin-orbit field correspond to particular cases in which the linear term in $q$ in Eq. (\ref{DL_strong_SOC}) vanishes.

When the field is longitudinal ($h_x = 0$), we get back to the FFLO case without SO, namely the modulated phase emerges at the tricritical point $(T^\star = 0.56\,T_{c0},\, h^\star = 1.07\,T_{c0})$. Indeed, when the spin-orbit and Zeeman fields are parallel, the SOC can be gauged out, and we obtain an effective Hamiltonian without SOC.

In the opposite case, namely when the Zeeman field is transverse to the SOC ($h_z = 0$), we obtain a rather different effect: The wave-vector $q$ vanishes, and $T_c$ is not modified with respect to $T_{c0}$ within leading order in $h_x/\alpha\,p_\text{F} \ll 1$. As we mentioned previously, this result comes from the fact that we neglect interband correlations in the self-consistency relation Eq. (\ref{app:SCE_fin}). 


We can also compare the $(h_x,\,h_z,\,T)$-phase diagrams for different values of the spin-orbit coupling constant $\alpha$ for $h_x = h_z$, as illustrated in Fig. \ref{diag_phaseSO}. The blue curve represents the transition line between the normal and superconducting states in the absence of SOC, whereas the three others correspond to three finite values of $\alpha$. Globally, the presence of spin-orbit interaction stabilizes the superconducting phase in a broader range of fields and temperature. Let us focus on the transition lines corresponding to $\alpha \neq 0$. At temperatures $T > T^\star = 0.56\,T_{c0}$, these curves cannot be distinguished from each other due to the very small values of $q$. These ones can be estimated from Eq. (\ref{q0_strong}) for a small Zeeman field: for example at $h = 0.5\,T_{c0}$, $\theta = \pi/
4$ and $\alpha = 0.05\,v_\text{F}$, $q \approx 10^{-10}\,p_\text{F}$, which weakly influences the critical temperature. But these values significantly increase for temperatures smaller than the tricritical temperature, related to the competition between helical and FFLO-like modulations \cite{agterberg_magnetoelectric_2012}: At large temperatures, the modulation coming from the helical state dominates, whereas below the tricritical temperature $T^\star = 0.56\,T_{c0}$, the modulation stems mainly from the FFLO-like state.

\section{\label{sec:level5} Conclusion}
We have studied the onset of nonuniform superconductivity in a one-dimensional wire in presence of Rashba spin-orbit coupling and Zeeman effect. The spin-orbit coupling stabilizes a FFLO-like 
modulated order parameter at low field and for all temperature below $T_{c0}$. We have derived the self-consistency relation for the modulated superconducting order parameter in two extreme regimes of weak and strong spin-orbit strength respectively. The analytical expressions for the wavevectors of the modulated state scale as $q \propto h_x^2\,h_z \alpha$ in the weak spin-orbit coupling regime ($\alpha p_\text{F} \ll T_c$), and 
$q \propto h_x^2\,h_z /\alpha$ in the strong $\alpha$ regime ($\alpha p_\text{F} \gg T_c$). When the ratio $\alpha\,p_\text{F}/h$ increases, superconducting correlations change from interband to intraband, which was also demonstrated in two dimensions in \cite{zwicknagl_critical_2017}. 

The Zeeman field-temperature phase diagrams have been obtained numerically using the self-consistency relations for arbitrary strength of the spin-orbit coupling. These phase diagrams exhibit anisotropy effects related to the angle between the Zeeman field and the wire axis. We have plotted the phase diagram in the case of strong spin-orbit coupling for different orientations of the Zeeman field and several values of the spin-orbit constant, and highlighted the anisotropy caused by SOC. At last we have shown that increasing $\alpha$ and the ratio $h_x/h_z$ allows superconductivity at larger temperature.

Finally, the results above also pertain for the case of quasi-one dimensional organic superconductors consisting in weakly coupled one-dimensional chains, although the spin-orbit should be due to the bulk structure of the crystal instead of the surface Rashba coupling. 

Some perspectives result from this work. Because of its sensitivity to impurities \cite{azlamazov_influence_1969}, it has been extremely difficult to observe experimentally the FFLO state. However, spin-orbit interaction should be sufficient to protect the modulated phase \cite{michaeli_superconducting_2012}. Thus one could emphasize that the nonuniform phase would be directly observable in an experimental system made of a superconductor coupled to a ferromagnet with strong spin-orbit interaction. 

Our model may be generalized to the two wire (or two plane) system, where superconductivity is generated in one wire (plane), while the spin-orbit and exchange interactions occur in the other one. Such model may be relevant for the description of S/F systems with spin-orbit interaction. Using a method similar to \cite{andreev_pi-phase_1991}, we should obtain the exact solution for this model in terms of the Gor'kov Green's functions. Using a SU(2) covariant method, the two-wire system has been studied in and shown to exhibit spontaneous currents \cite{Baumard_2020}. 

Moreover, we outline that the system described in the present paper could be used as a link between two identical superconductors to create an anomalous Josephson junction \cite{krive_influence_2005,buzdin_direct_2008}, similarly as \cite{szombati_josephson_2016}. Note that in \cite{martin_majorana_2012} it was demonstrated that a superconductor with a conical helical magnetic structure is described by the same model as a 1D wire with spin-orbit and exchange interactions. The theoretical analysis of the Josephson effect in 1D conical ferromagnet \cite{meng_nonuniform_2019} demonstrates the emergence of the $\varphi_0$-junction. In such a system, the modulation parameter $q$ would play the role of the phase difference needed to generate a current in the junction, which would open new possibilities of application of these $\varphi_0$-junctions in memory devices \cite{padurariu_theoretical_2010}.

\section{Acknowledgements}
The authors thank F. S. Bergeret for useful discussions and fruitful remarks and suggestions. This work was supported by EU Network COST CA16218 (NANOCOHYBRI) and the French ANR project OPTOFLUXONICS (A.B. and J.C.). J. B. acknowledges the financial support from the Initiative d'Excellence (IDEX) of the Universit\'e de Bordeaux and the funding by the Spanish Ministerio de Econom\'ia y Competitividad (MINECO) (Projects No. FIS2014-55987-P and No. FIS2017-82804-P).  
Computer time for this study was provided by the computing facilities MCIA (M\'esocentre de Calcul Intensif Aquitain) of the Universit\'e de Bordeaux and of the Universit\'e de Pau et des Pays de l'Adour.\\

\medskip 

{\bf \large Author contribution statement}

\medskip

All the authors have contributed to the derivation of the analytical results.
Moreover Julie Baumard has also implemented the numerical calculations for the various phase diagrams.
All the authors have contributed to the redaction of the manuscript and approved its final version.

\appendix
\section{\label{app:hamiltonian} Normal state Hamiltonian in the helical basis}
Here we provide details concerning the derivation of the diagonal Hamiltonian Eq. (\ref{hamiltonian_diag}). In Nambu space, the Hamiltonian describing the wire in the absence of superconductivity may be written:
\begin{equation}
    \hat{h}_\text{N} = \left(\xi + \alpha\,p\,\sigma_z\right)\,\tau_z + \vec{h}\cdot\vec{\sigma} \;,
\end{equation}
where $\sigma_{x,y,z}$ and $\tau_{x,y,z}$ are the Pauli matrices acting in spin and Nambu space respectively.

The idea is to rotate the spin Pauli matrices $\sigma_x$ and $\sigma_z$ around $\sigma_y$, by a suitable angle $\phi_p$ :
\begin{equation}
\phi_p = \displaystyle \arccos\left(\frac{h_z + \alpha\,p}{\sqrt{h_x^2 + \left(h_z + \alpha\,p\right)^2}}\right) \, ,
\end{equation}
in the $\tau_z = +1$ sector, and by the distinct angle $(\pi + \phi_{-p})$ within the $\tau_z = -1$ (hole) sector.
For this purpose, we introduce the matrix :
\begin{equation}
\mathcal{U} = \begin{pmatrix}
\tilde{\mathcal{U}}\left(\phi_p\right) & 0 \\
0 & \tilde{\mathcal{U}}\left(\pi + \phi_{-p}\right)
\end{pmatrix} \;, 
\end{equation}
where $\tilde{\mathcal{U}}\left(\phi_p\right) = e^{-\text{i}\frac{\phi_p}{2}\sigma_y}$ is the spin rotation matrix in the $\tau =+1$ sector. 

Using the transformation $\hat{h}_\text{N} \rightarrow \mathcal{U}^\dagger\,\hat{h}_\text{N}\,\mathcal{U} = \hat{h}_\text{hel}$, one can write the Hamiltonian operator in the helical basis:
\begin{equation}
    \hat{h}_\text{hel} = \xi\,\tau_z - \frac{\lambda}{2}\left[\left(\varepsilon_+ + \varepsilon_-\right)\,\tau_0 + \left(\varepsilon_+ - \varepsilon_-\right)\,\tau_z\right]\;,
\end{equation}
where $\varepsilon_\pm = \sqrt{h_x^2 + \left(h_z \pm \alpha\,p\right)^2}$ and $\lambda = \pm$ labels the energy bands. 

\section{\label{app:Adding_SC} Superconductivity in the helical basis}
Since the system is studied near the normal/superconducting transition ($\Delta_0 \ll T_c$), superconductivity can be treated perturbatively. To take into account the momentum shift $q$ caused by the modulation, it is useful to define the "shifted quantities", labelled by the superscript $q$ as $\mathcal{U} \rightarrow \mathcal{U}^{(q)}$ and $\hat{G}_0 \rightarrow \hat{G}_0^{(q)}$, with
\begin{equation}
\mathcal{U}^{(q)} = \begin{pmatrix}
\tilde{\mathcal{U}}\left(\phi_{p + \frac{q}{2}}\right) & 0 \\
0 & \tilde{\mathcal{U}}\left(\pi + \phi_{-p + \frac{q}{2}}\right)
\end{pmatrix} \;,
\label{Uq}
\end{equation}
and 
\begin{equation}
    \hat{G}_0^{(q)} = \begin{pmatrix}
    G_\lambda\left(p + \frac{q}{2},\, \omega_n\right) & 0\\
    0 & -G_\lambda\left(-p + \frac{q}{2},\, -\omega_n\right)
    \end{pmatrix} \;.
\end{equation}
Then, to first order in $\Delta_0$, the Green's function matrix in the presence of superconductivity is:
\begin{equation}
    \hat{G}^S = \hat{G}_0^{(q)} - \Delta_0\hat{G}_0^{(q)} \left(\mathcal{U}^{(q)}\right)^\dagger\tau_x\,\mathcal{U}^{(q)}\,\hat{G}_0^{(q)} \;.
\end{equation}

\medskip

\section{\label{app:SCE} Derivation of the self-consistency relation}
In the helical basis, the self-consistency equation may be written
\begin{equation}
    \Delta_0 = \frac{1}{4}\,|\gamma|\,T\,\sum_{\omega_n} \int_{-\infty}^{+\infty} \text{Tr}\left(\hat{G}^S\left(\mathcal{U}^S\right)^\dagger\tau_x\,\mathcal{U}^S\right) \frac{\mathrm{d}p}{2\,\pi} \;.
    \label{app:eq_SCE1}
\end{equation}
The trace in Eq. (\ref{app:eq_SCE1}) contains two finite terms:
\begin{align}
    &\text{Tr}\left(\hat{G}^S\left(\mathcal{U}^S\right)^\dagger\tau_x\,\mathcal{U}^S\right) = \notag\\  
    &2\,\Delta_0\left[\sin^2\left(\frac{\eta}{2}\right)\sum_\lambda G_\lambda\left(p + \frac{q}{2},\, \omega_n\right)\,G_\lambda\left(-p + \frac{q}{2},\, -\omega_n\right) \right.\notag \\ 
    & \left. + \cos^2\left(\frac{\eta}{2}\right) \sum_{\lambda,\lambda^\prime} G_\lambda\left(p + \frac{q}{2},\, \omega_n\right)\,G_{\lambda^\prime}\left(-p + \frac{q}{2},\, -\omega_n\right)\right] \;,
    \label{Trace}
\end{align}
where $\eta = \phi_{-p + q/2} - \phi_{p + q/2}$. The first term of Eq. (\ref{Trace}) describes intraband pairing correlations, whereas the second one corresponds to interband pairing. Using this expression, the self-consistency relation can be put in a more convenient form, thus eliminating $\gamma$:
\begin{equation}
    \ln\left(\frac{T_c}{T_{c0}}\right) = 2\,T_c\,\sum_{\omega_n \geq 0} \left[\frac{v_\text{F}}{8}\,\text{Re}\left(\int_{-\infty}^{+\infty} \text{Tr}\left(\tilde{F}\right)\mathrm{d}p\right) - \frac{\pi}{\omega_n}\right]\;,
    \label{app:eq_SCE2}
\end{equation}
where $\tilde{F} = -\left(\hat{G}^S\left(\mathcal{U}^S\right)^\dagger\tau_x\,\mathcal{U}^S\right)/\Delta_0$. 

In the regime of strong spin-orbit interaction, $\alpha\,p_\text{F} \gg h_x,\,h_z,\,T_c$ and $E_\text{so} = \frac{1}{2}\,m\,\alpha^2 \ll \mu$, the phase difference $\eta \rightarrow \pi$ at zeroth order in the Zeeman field, thus allowing to neglect interband pairing terms. Therefore, the self-consistency relation Eq. (\ref{app:eq_SCE2}) becomes :
\begin{widetext}
\begin{equation}
    \ln\left(\frac{T_c}{T_{c0}}\right) = T_c\,\sum_{\omega_n \geq 0,\,\lambda} \left[\frac{v_\text{F}}{2}\,\text{Re}\left(\int_{-\infty}^{+\infty} \mathcal{G}_\lambda\left(p + \frac{q}{2},\, \omega_n\right)\,\mathcal{G}_\lambda\left(-p + \frac{q}{2},\, -\omega_n\right)\mathrm{d}p\right) - \frac{\pi}{\omega_n}\right]\; .
    \label{app:SCE_helical_intrabande}
\end{equation}
\end{widetext}

The method used to perform the $p$ integration in Eq. (\ref{app:SCE_helical_intrabande}) is quite different from the one used in Sec. \ref{subsec:DL_weak} for the small spin-orbit limit. Indeed in the regime of strong spin-orbit interaction, we consider only intraband pairing terms, which allows to treat each band separately. Therefore, it is possible to compute the integral in the customary way \cite{dimitrova_two-dimensional_2006}. Because  the Green's functions $\mathcal{G}_\lambda$ are peaked around the Fermi energy $p_\text{F}^\lambda$, we can turn the $p$ integral into a $\xi$ integral but for each band separately:
\begin{equation}
    \int \mathrm{d}p \rightarrow \frac{1}{v_\text{F}}\int \mathrm{d}\xi \;,
    \label{app:int_xi_helical}
\end{equation}
and we can approximate the momentum $p$ by $p_\text{F}^\lambda$ in the $\alpha\,p$ and $q\,p$ terms, where
\begin{equation}
    p_\text{F}^\lambda = \lambda\,m\,\alpha + \sqrt{2\,m\,\mu + m^2\,\alpha^2} \;,
    \label{app:p_lambda}
\end{equation}
is the Fermi momentum of the band $\lambda$, obtained by neglecting the Zeeman contribution with respect to the SOC: $h_x,\,h_z \ll \alpha\,p_\text{F}$. In such a way  the $\xi$ integration can be  performed straightforwardly.
\begin{widetext}
\begin{equation}
    \ln\left(\frac{T_c}{T_{c0}}\right) = 2\,\pi\,T_c\,\sum_{\omega_n \geq 0,\,\lambda}\left[\frac{2\,\omega_n}{4\,\omega_n^2 + \left[q\,v_\text{F}^\lambda - \lambda\left(\sqrt{h_x^2 + \left(h_z + \alpha\,p_\text{F}^\lambda + \alpha\,q/2\right)^2} - \sqrt{h_x^2 + \left(h_z - \alpha\,p_\text{F}^\lambda + \alpha\,q/2\right)^2}\right)\right]^2}  - \frac{1}{2\,\omega_n}\right] \;.
    \label{app:SCE_fin}
\end{equation}
\end{widetext}

\bibstyle{apsrev4-1}
\bibliography{References}

\end{document}